\documentclass{aa}  

\usepackage{graphicx}
\usepackage{amssymb,amsmath}
\usepackage{latexsym}
\usepackage{hyperref}								
\usepackage{lmodern}
\usepackage{mathtools}
\usepackage{epsfig}
\usepackage{color}
\usepackage{cancel}
\usepackage[varg]{txfonts}

\makeatletter
\renewcommand*\aa@pageof{, page \thepage{} of \pageref*{LastPage}}
\makeatother

\def\kk{\mathtt{k}}
\def\H{\mathcal{H}}

\begin{document}
\title{Maps of secular resonances in the NEO region}
\titlerunning{Secular resonances in the NEO region}

   \author{Marco Fenucci\inst{1,2,3}
          \and
          Giovanni F. Gronchi\inst{4}
          \and
          Bojan Novakovi\'c\inst{1}
          }
   \authorrunning{M. Fenucci et al.}
   \institute{Department of Astronomy, Faculty of Mathematics, University of Belgrade,
             Studentski trg 16, 11000 Belgrade, Serbia\\
               \email{marco.fenucci@ext.esa.int}
               \and
               ESA NEO Coordination Centre, Largo Galileo Galilei, 1, 00044 Frascati, Italy
               \and
               Elecnor Deimos, Via Giuseppe Verdi, 6, 28060 San Pietro Mosezzo, Italy
               \and
               Dipartimento di Matematica, Università di Pisa, Largo  B. Pontecorvo 5, 56127 Pisa, Italy
             }

   \date{Received --- / Accepted ---}


  \abstract
     {From numerical simulations, it is known that some secular resonances may affect the
        motion of near-Earth objects (NEOs). However, the specific location of the secular
        resonance inside the NEO region is not fully known, because the methods previously used to
        predict their location can not be used for highly eccentric orbits and when the NEOs cross the orbits of the planets.}
     {In this paper, we aim to map the secular resonances with the planets
     from Venus to Saturn in the NEO region, even for high values of the eccentricity.}
     {We used an averaged semi-analytical model that can deal with orbit crossing singularities for the
     computation of the secular dynamics of NEOs, from which we can obtain suitable
  proper elements and proper frequencies. Then, we computed the proper frequencies over a uniform grid in the
  proper elements space. Secular resonances are thus located by the level curves
  corresponding to the proper frequencies of the planets.}
     {We determined the location of the secular resonances with the planets from Venus to
     Saturn, showing that they appear well inside the NEO region. By using full numerical
  $N$-body simulations we also showed that the location predicted by our method is
  fairly accurate. Finally, we provided some indications about possible dynamical paths inside the NEO
  region, due to the presence of secular resonances.}
     {}



   \maketitle

\section{Introduction}
In the dynamics of the $N$-body problem, with $N>2$, the geometry of the orbits of the bodies
changes with time due to their mutual gravitational interactions. In
particular, the angles that specify the orientation of the osculating ellipse precess at a secular rate. 
In the context of our Solar System, the precession frequencies of the longitude of the perihelion
$\varpi$ and of the longitude of the ascending node $\Omega$ of the orbits of the planets are typically denoted with $g_j$ and $s_j$, respectively, 
where the index $j=1,\dots,8$ labels the planets from Mercury to Neptune. Accurate values of planetary secular
frequencies were computed, for instance, by \citet{nobili-etal_1989} and \citet{laskar-etal_2004,
laskar-etal_2011}.

The presence of the planets also cause both short and long-term perturbations on the
motion of small Solar System bodies, such as asteroids. The amplitude of the secular
perturbations increases near a \textit{secular resonance}, that occurs when the
precession frequency $g$ of the longitude of the perihelion, or the frequency $s$ of the
longitude of the node, of the
asteroid is nearly equal to the one of a planet.
Linear secular resonances involve only one asteroid's frequency and one planet's
frequency, and resonances of type $g\simeq g_j$ are typically denoted with $\nu_j$, while those of type
$s \simeq s_j$ are denoted with $\nu_{1j}$.

The problem of locating the secular resonances in the main belt underwent a series of
improvements, and it is nowadays a well known topic in asteroid dynamics. The first
attempt to produce maps of secular resonances in the main belt was done by
\citet{williams_1969}, who used a semi-analytical method to derive asteroids' proper
elements and proper frequencies.
The method was later refined by \citet{williams-faulkner_1981} to better locate the three
strongest secular resonances, namely $\nu_5, \nu_6$, and $\nu_{16}$.
Later on, a simple analytical model was used by \citet{yoshikawa_1987} to locate more precisely the
$\nu_6$ secular resonance.
More accurate analytical theories, where the perturbing function is expanded in Fourier series of cosines of combinations of the angular variables and the amplitudes are expressed as polynomials in eccentricity and sine of inclination, were developed by \citet{milani-knezevic_1990,
milani-knezevic_1994}. The authors were able to produce maps of both linear and non-linear
secular resonances in the main belt, for small proper eccentricity and proper inclination.
The same method was used by \citet{knezevic-etal_1991} to map the location of all linear
secular resonances with the outer planets, for heliocentric distances between 2 and 50 au. 
A semi-analytical theory that avoids the expansion in eccentricity and inclination has been
developed by \citet{morbidelli-henrard_1991, morbidelli-henrard_1991b}, and it permitted
the authors to locate the linear secular resonances at higher values of eccentricity and
inclination.
More recently, the introduction of a synthetic theory for the computation of proper
elements and frequencies \citep[see][]{knezevic-milani_2000, knezevic-milani_2003} led to
attempt to locate secular resonances with purely numerical techniques.  A first empirical
method was used by \citet{tsirvoulis-novakovic_2016}, who identified a large population of
synthetic asteroids in a secular resonance with (1) Ceres and (4) Vesta by matching their
secular frequencies up to a certain threshold.  A new, completely synthetic method was
recently developed by \citet{knezevic-milani_2019}. This method was used by
\citet{knezevic_2020} to significantly improve the location of the $\nu_5$ resonance, and
by \cite{knezevic_2022} to review the location of the linear and several non-linear
secular resonances in the main belt.

On the contrary, little attention has been paid to the location of secular
resonances in the region of near-Earth objects (NEOs), that are asteroids with a
perihelion distance $q$ smaller than 1.3 au.  So far, the first and only attempt was done 
by \citet{michel-froeschle_1997}, who used a semi-analytic model similar to that of
\citet{kozai_1962} to locate the linear secular resonances at semi-major axis $a$ between
0.5 and 2 au, only for a value of proper eccentricity equal to 0.1. However, NEOs can
reach much higher eccentricities, and they may cross the orbit of one or more planets. Orbit
crossings were indeed the reason for which \citet{michel-froeschle_1997} were not able to
compute maps at higher proper eccentricities. 
In turn, locating the secular resonances inside the NEO region may help to better
understand the global dynamics of NEOs. From purely numerical simulations, it is now known
that secular resonances play a fundamental role in the delivery of objects from the
main belt to the NEO region \citep[see, e.g.,][]{scholl-froeschle_1986,
froeschle-scholl_1986, froeschle-scholl_1989, granvik-etal_2017}. In particular, the $\nu_6$ secular
resonance is the most important mechanism for the production of NEOs, and it provides the
largest fraction of Earth's impactor \citep{bottke-etal_2002, granvik-etal_2018}. Although
planetary close encounters are one of the main perturbations in the motion of NEOs,
secular resonances may still play a significant role in their dynamics
\citep{foschini-etal_2000, gladman-etal_2000}.

In this work, we located the linear secular resonances with the planets from Venus to
Saturn inside the NEO region, for semi-major axis between 0.5 and 3 au.  For this purpose,
we used the semi-analytic method by \citet{gronchi-milani_2001} to propagate the secular
dynamics of NEOs, from which the proper frequencies $g$ and $s$ are extracted.  Proper
frequencies are then computed on a grid of points in the phase spaces of $(a, e_{0})$ or
$(a,i_{0})$, where $e_0$ are $i_0$ are the proper eccentricity and proper inclination.
Maps of the secular resonances with planets from Venus to Saturn were then determined by
computing the level curves corresponding to planets' proper frequencies.  The method by
\citet{gronchi-milani_2001} permits us to compute the secular evolution even beyond an
orbit crossing, thus enabling us to produce secular resonance maps even at high
eccentricity.

The paper is structured as follows. In Section~\ref{s:methods} we introduce the
semi-analytical Hamiltonian model for the computation of the secular evolution of NEOs, 
and we describe the method to locate the secular resonances. In Section~\ref{s:results} we 
show the location of the secular resonances obtained with our method, and show some
comparison with full numerical simulations. In Section~\ref{s:discussion}, we reported the
implication of our results on the global dynamics of NEOs, and we discussed the effects of
mean-motion resonances. Finally, in Section~\ref{s:conclusions} we provide our
conclusions.

\section{Methods}
\label{s:methods}

\subsection{The Hamiltonian secular model}
We used the averaged model by \citet{gronchi-milani_2001} to compute the secular evolution of an asteroid. The planets from Venus to Saturn are included in the model, and they are placed on circular co-planar orbits.
We denote the Gauss constant with $\kk = \sqrt{\mathcal{G}m_0}$, where
$\mathcal{G}$ is the universal gravitational constant, and $m_0$ is the mass of the Sun.
We also set $\mu_j = m_j/m_0, \ j=2,\dots,6$, where $m_j$ is the mass of the $j$-th
planet. 
The canonical Delaunay variables are defined as
\begin{equation}
\begin{split}
    L = \kk\sqrt{a},              & \qquad \ell = M, \\
    G = \kk\sqrt{a(1-e^2)},       & \qquad    g = \omega, \\
    Z = \kk\sqrt{a(1-e^2)}\cos i, & \qquad    z = \Omega,
\end{split}
\label{eq:delaunay}
\end{equation}
where $a$ is the semi-major axis, $e$ is the eccentricity, $i$ is the inclination, $M$ is
the mean anomaly, $\omega$ is the argument of the pericenter\footnote{Note that the symbol $g$ is also used to denote the secular frequency of the longitude of perihelion $\varpi$, as mentioned in the Introduction. However, it is clear from the context when we use $g$ to refer to the argument of the pericenter $\omega$, or to refer to the frequency of $\varpi$.}, and $\Omega$ is the
longitude of the ascending node. The Hamiltonian describing the motion of the asteroid is
\begin{equation}
   \mathcal{H} = \mathcal{H}_0 + \varepsilon \mathcal{H}_1, \qquad \varepsilon=\mu_5,
   \label{eq:hamiltonian}
\end{equation}
where $\mathcal{H}_0$ is the unperturbed Keplerian part, and $\mathcal{H}_1$ accounts for
the perturbations of the planets:
\begin{equation}
    \begin{split}
        \H_0 & = -\frac{\kk^4}{2L^2}, \\
        \H_1 & = -\kk^2\sum_{j=2}^6  \frac{\mu_j}{\mu_5}
        \bigg( 
        \frac{1}{|\mathbf{r}-\mathbf{r}_j|} - \frac{\mathbf{r} \cdot \mathbf{r}_j}{|\mathbf{r}_j|^3}
        \bigg).
    \end{split}
    \label{eq:hamiltonianTerms}
\end{equation}
In Eq.~\eqref{eq:hamiltonianTerms}, $\mathbf{r}$ and $\mathbf{r}_j$ denote the
heliocentric position of the asteroid and the $j$-th planet, respectively.

Assuming that there are no mean-motion resonances between the asteroid and any of the
planets, the mean anomalies $\ell$ and $\ell_j$ of the asteroid and the $j$-th planet
evolve much faster than the other orbital elements. The secular evolution at
first order in the planetary masses is then obtained by the double average of the
Hamiltonian of Eq.~\eqref{eq:hamiltonian} over the fast angles, i.e.
\begin{equation}
   \overline{\H} = -\varepsilon \kk^2
   \sum_{j=2}^6  \frac{\mu_j}{\mu_5}\int_0^{2\pi}\int_{0}^{2\pi}
        \frac{1}{|\mathbf{r}-\mathbf{r}_j|}
        \text{d}\ell\text{d}\ell_j.
   \label{eq:averHamiltonian}
\end{equation}
With the assumptions of zero eccentricity and inclination for the orbits of the planets, this model can be seen as the dominant term of an expansion in powers of the planetary eccentricities and inclinations of the double-averaged Hamiltonian in the case of eccentric and inclined orbits of the planets \citep{thomas-morbidelli_1996}.
Note that the indirect perturbation is not present in the averaged Hamiltonian
$\overline{\H}$, since its double integral vanishes. In this model, $L$ is a first integral,
hence the semi-major axis $a$ remains constant. Moreover, $\Omega$ does not appear
in the Hamiltonian of Eq.~\eqref{eq:averHamiltonian} because the planets are placed on
circular and co-planar orbits, and the conjugate momentum $Z$ is also a first integral. The system
defined by $\overline{\H}$ has therefore only one-degree of freedom, and it is integrable. 
Delaunay elements are classically used to study this problem \citep[see e.g.][]{kozai_1962, thomas-morbidelli_1996, gallardo-etal_2012, saillenfest-etal_2016}, and this is because the Hamiltonian in these coordinates can be naturally reduced to a 1-degree of freedom Hamiltonian depending only on $(G, g)$.

The averaged Hamiltonian $\overline{\H}$ has a first
order polar singularity when the orbit of the asteroid crosses that of a planet
\citep{gronchi-milani_1998}. By denoting with $\smash{y \in \{G,Z,g,z\}}$ one of the Delaunay elements, 
the derivatives
$\partial\overline{\H}/\partial y$ can be computed numerically simply by exchanging
the derivative and the integral signs when there are no orbit crossings. On the other hand, at orbit crossings the double
integral of the derivatives of $\H_1$ is divergent, and therefore the
theorem of differentiation under the integral sign does not apply.
\citet{gronchi-milani_1998, gronchi-tardioli_2013} showed that, in a neighbourhood of an orbit crossing
configuration, the vector field defined by the averaged Hamiltonian has a two-fold
definition, that can be computed analytically. 
By using this property, \citet{gronchi-milani_2001} set up a numerically stable algorithm
that permits to compute a solution of Hamilton's equations for $\overline{\H}$ beyond an orbit crossing
configuration. In this work we used this algorithm to propagate the secular evolution of an asteroid beyond an orbit crossing singularity.
Details about the method and the practical numerical implementation can be found in \citet{gronchi-milani_1998, gronchi-milani_2001, gronchi-tardioli_2013, fenucci-etal_2022}.

\subsection{Locating the secular resonances}
Proper frequencies can be computed as in \citet{gronchi-milani_2001}. A solution of the
Hamiltonian secular system defined by $\overline{\H}$ is 
computed numerically and, if $\omega$ circulates, we can find the times $t_0$ and 
$t_{\pi/2}$ such that $\omega(t_0) = 0, \ \omega(t_{\pi/2}) = \pi/2$. We also set 
$\Omega_0 = \Omega(t_0), \ \Omega_{\pi/2} = \Omega(t_{\pi/2})$. 
The Hamiltonian $\overline{\mathcal{H}}$ is even and $\pi$-periodic in $\omega$, therefore the frequency $g-s$ of the argument of perihelion $\omega$ is given by
\begin{equation}
   g-s = \frac{2\pi}{4(t_{\pi/2} - t_0)}, 
   \label{eq:gms}
\end{equation}
while the frequency $s$ of the longitude of the node $\Omega$ is given by
\begin{equation}
   s = \frac{\Omega_{\pi/2} - \Omega_0}{t_{\pi/2} - t_0}.
   \label{eq:s}
\end{equation}
The dynamical state in which $\omega$ librates is also called Kozai resonance. In this work we assume that $\omega$ librates around $0^\circ$, therefore we take
$t_0$ such that $\omega(t_0)=0$ and $\dot{\omega}(t_0)>0$. Then, we compute the time
$\Delta t$ after which $\omega$ vanishes again
and we denote
by $\Delta \Omega$ the change in $\Omega$ during this timespan. The frequency $g-s$ of
$\omega$ vanishes, but we can still define a libration frequency $\text{lf}$, and the
frequency $s$ for $\Omega$, as
\begin{equation}
   \text{lf} = \pm\frac{2\pi}{2\Delta t}, \quad s= \frac{\Delta \Omega}{\Delta t}.
   \label{eq:lf}
\end{equation}
The formula for the libration frequency lf is justified by the symmetry properties of $\overline{\mathcal{H}}$. 
Besides the proper frequencies, the proper eccentricity and the proper inclination can be
defined as $e_0 = e(t_0)$ and $i_0 = i(t_0)$, respectively. 
This is a slightly different definition from that adopted by \citet{gronchi-milani_2001}, who used the minimum
eccentricity $e_{\min}$, and the corresponding maximum inclination $i_{\max}$, attained along a solution. 
However, the two definitions coincide 
in the case that the
solution does not cross any planet's orbit \citep{michel-froeschle_1997}.
Note that $\omega$ may also librate around other values, and therefore it may never vanish during a libration period \citep[see, e.g.][]{michel-froeschle_1997, gronchi-milani_1998, gronchi-milani_2001}. While these regions certainly exist, they do not appear in our results, because they would need another special definition of the proper elements based on the center of their libration island.

The method to identify the secular resonances is similar to that of
\citet{michel-froeschle_1997}. The initial longitude of the node $\Omega$ is set to 0, but any
other value would not change the result since $\Omega$ does not appear in the Hamiltonian
$\overline{\H}$. The initial argument of pericenter $\omega$ is also set to 0, so that the
initial eccentricity and inclination also correspond to the proper values.
Then, we fix the initial value of the proper eccentricity $e_{0}$ (or the initial proper
inclination $i_{0}$), and choose a grid in the $(a, i_{0})$-plane (or in the
$(a,e_{0})$-plane). 
The proper frequencies $g$ and $s$ are computed for each point of the grid.
The location of the secular resonances $\nu_j, \, \nu_{1j}, \ j=2,\dots,6$,
in the $(a,i_{0})$-plane or in the $(a,e_{0})$-plane, is then given by
the contour lines corresponding to the levels $g=g_j$ and  $s=s_j, \ j=2,\dots,6$, respectively.
In this work, we used the values of the secular frequencies of the planets determined by
\citet{laskar-etal_2011}, that are reported in Table~\ref{tab:planet_frequencies}.

\begin{table}
   \renewcommand{\arraystretch}{1.3}
   \centering
      \caption{Secular frequencies of the planets from Venus to Saturn, as determined by
      \citet{laskar-etal_2011}.}
   \begin{tabular}{ccc}
      \hline
      \hline
      Planet & $g$ ($^{\prime\prime}/$yr) & $s$ ($^{\prime\prime}/$yr) \\
      \hline
      Venus   & 7.453    & -7.06      \\
      Earth   & 17.368   & -18.848    \\
      Mars    & 17.916   & -17.751    \\
      Jupiter & 4.257482 & 0.0        \\
      Saturn  & 28.2449  & -26.347841 \\
      \hline
   \end{tabular}
   \label{tab:planet_frequencies}
\end{table}

\section{Results}
\label{s:results}

\subsection{Secular resonance maps at fixed inclination}
\label{ss:map_ae}
\begin{figure*}[!ht]
   \centering
   \includegraphics[width=\textwidth]{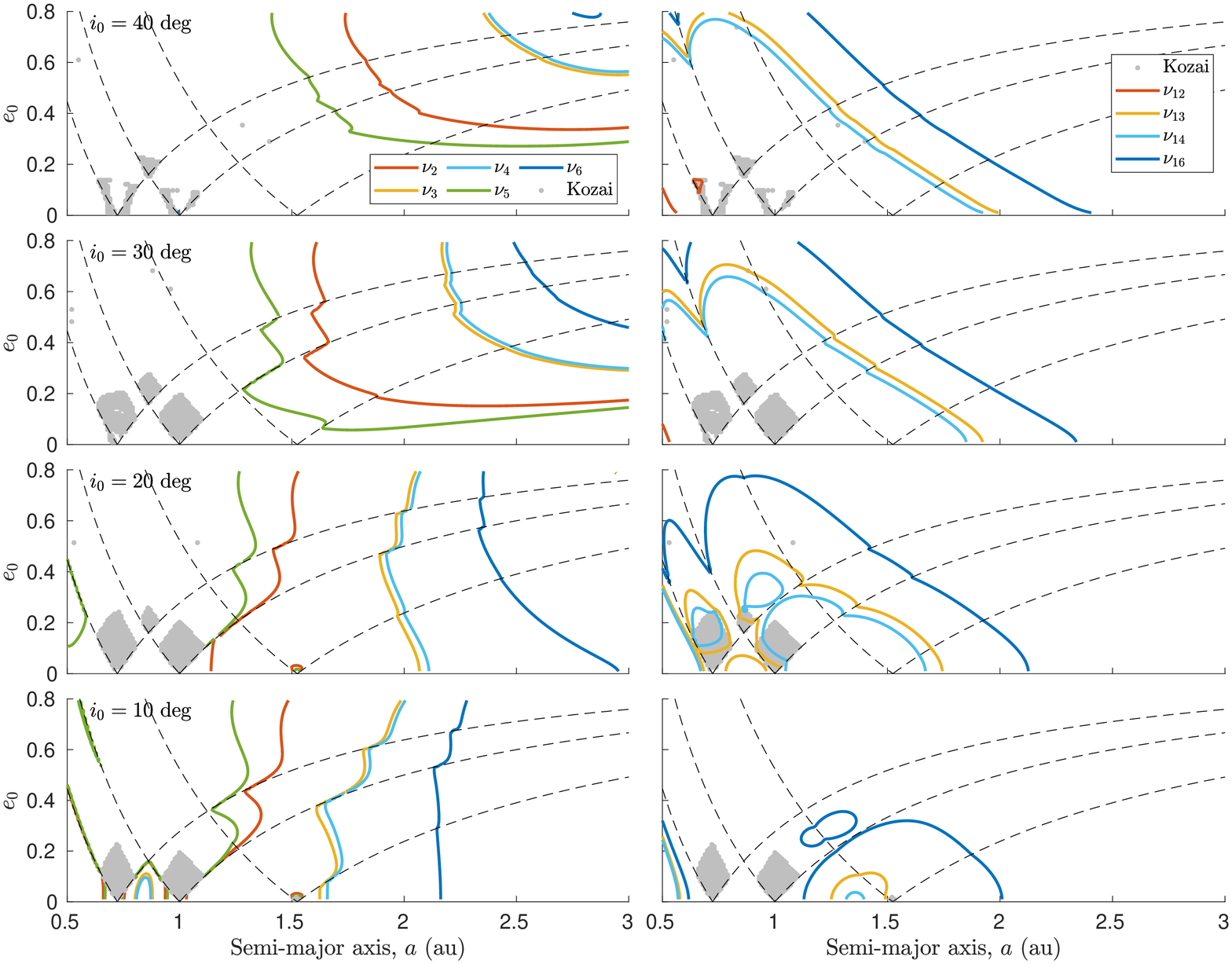}
   \caption{Map of secular resonances in the $(a,e_{0})$-plane for $a$ between 0.5 and 3 au, for different values of
      proper inclination $i_{0}$. The left column shows the locations of $\nu_j, j=2,3,4,5,6$, while the
      right column shows the locations of $\nu_{1j}, j=2,3,4,6$. 
      The color code of each resonance is indicated by the legend in the top row panels.
      From the top to the bottom, rows correspond to values of $i_{0}$ equal to $40^\circ,
      30^\circ, 20^\circ$, and $10^\circ$, respectively.
      The dashed curves correspond to $q=a_{2}, a_{3},
      a_{4}$ and $Q=a_{2}, a_{3},
      a_{4}$, where $q=a(1-e), Q = a(1+e)$ are the perihelion and the aphelion
      distance of the asteroid, respectively. We also reported the location of the Kozai
      librators in grey, identified by $g -s= 0$.
   }
   \label{fig:map_ae_big}
\end{figure*}
\begin{figure*}[!ht]
   \centering
   \includegraphics[width=\textwidth]{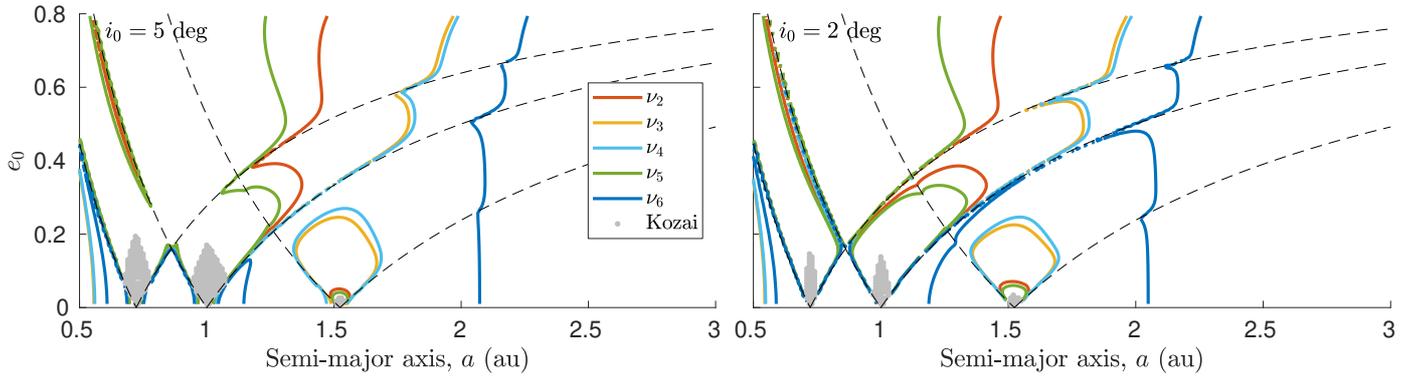}
   \caption{Map of the $\nu_j, j=2,3,4,5,6$ secular resonances in the $(a,e_{0})$-plane
   for $a$ between 0.5 and 3 au, at $i_{0}=5^\circ$
   (left panel), and at $i_{0}=2^\circ$ (right panel). Dashed curves correspond to
   $q=a_{2}, a_{3},
      a_{4}$ and $Q=a_{2}, a_{3},
      a_{4}$.  We also reported the location of the Kozai
      librators in grey, identified by $g -s= 0$.}
   \label{fig:map_ae_small}
\end{figure*}
We discretized the $(a,e_{0})$-plane with a step of 0.01 au for the semi-major axis, and
0.008 for the proper eccentricity. We took into account the region defined by 0.5 au $<a<$ 3 au 
and $0.01 < e_{0} < 0.8$. The proper inclination was fixed, and we computed the maps
for $i_{0} = 2^\circ,5^\circ,10^\circ,20^\circ,30^\circ,$ and $40^\circ$.

The left column of Fig.~\ref{fig:map_ae_big} shows the location of the $\nu_j, j=2,\dots,6$ secular
resonances, while the right column shows that of $\nu_{1j}, j=2,3,4,6$, for values of
proper inclination $i_{0}$ equal to $40^\circ, 30^\circ, 20^\circ$, and $10^\circ$. 
Figure~\ref{fig:map_ae_small} shows the location of the $\nu_j, j=2,\dots,6$ resonances at lower
inclination, namely for $i_{0} = 5^{\circ}, 2^{\circ}$. 
Figure~\ref{fig:map_ae_big} and \ref{fig:map_ae_small} also show the
curves $\smash{q = a_{2}, a_{3}, a_{4}}$ and 
$Q = a_{2}, a_{3}, a_{4}$, where $q=a(1-e)$ and $Q=a(1+e)$
are the perihelion and the aphelion distance of the asteroid, respectively, and $a_j$ denote the semi-major axis of the planets.

From the left column of Fig.~\ref{fig:map_ae_big} we can see that $\nu_j, \, j=2,3,4,5$
(i.e. resonances with the inner planets, and with Jupiter) always appear inside the NEO
region, while the $\nu_6$ secular resonance with Saturn appears to a large extent at
proper inclinations $i_0$ smaller than $30^\circ$.
The resonances $\nu_3$ with the Earth and $\nu_4$ with Mars are always close to each
other, because the precession rates of the longitude of the perihelion of these two planets
are similar (see Table~\ref{tab:planet_frequencies}).
At $i_0 = 40^\circ$, all the secular resonances are confined at values of eccentricity
$e_0$ larger than 0.2, and at semi-major axis $a$ larger than about $1.3$ au.
As the proper inclination $i_0$ decreases, the secular resonances move towards smaller
values of semi-major axis, and they span an increasing interval of eccentricity. 
It is worth noting that, when passing from $i_0 = 30^\circ$ to $i_0 = 20^\circ$, there is
a significant change in the location of the $\nu_2$ and $\nu_5$ resonances. For inclination larger than
$30^\circ$ these two resonances are present in the region $q<a_{4}$, and they
are confined at rather low eccentricity, while for $q>a_{4}$ they extend towards
larger eccentricity values, up to the limit of $0.8$ we considered.  For inclination smaller than
$30^\circ$ they disappear from the region $q<a_{4}$, and they remain confined
at semi-major axis smaller than 1.5 au, where they assume values of eccentricity from near
0 up to 0.8.
The remaining resonances do not show this change for inclination between $10^\circ$ and $40^\circ$.
Additionally, the $\nu_5$ resonance shows a second branch at $a<1$ au for $i_0=20^\circ$.
The same happens to $\nu_3$ and $\nu_4$ for $i_0=10^\circ$.
%

The secular resonance maps for $\smash{i_0 = 2^\circ}$ and 5$^\circ$ (see
Fig.~\ref{fig:map_ae_small}) become more complex. At $i_0 = 5^\circ$, the $\nu_3$ and
$\nu_4$ disappear from the region $q<a_{4}$, while a new almost-circular branch
appears in the region $a_{3} < q < a_{4}$. In the region $q>a_{3}$,
these two resonances are close to the curve $q=a_{3}$ for semi-major axis smaller
than about 1.7 au, at which they start extending towards larger values of eccentricity. A
similar feature is seen for $\nu_2$ and $\nu_5$, that tend to follow the curves $q =
a_{3}$ and $q= a_{2}$ for certain intervals of semi-major axis
values. 
On the contrary, the $\nu_6$ resonance does not change much with respect to the one
obtained at $\smash{i_0 = 10^\circ}$.
Note also that more branches of all the resonances appear at semi-major axis smaller than
about 1.2 au, and most of them follow the curves of constant $Q$. 
The global picture for $i_0 = 2^\circ$ is similar to the previous case, with the
difference that the alignment along the curves $q=a_{2}$ and $q=a_{3}$ is
more definite, and it appears also for the $\nu_6$ secular resonance. 
These results show that, when the orbital elements get closer to the curves of perihelion
distance (or, to a minor extent, aphelion distance) equal to that of a planet, the value of
the proper frequency $g$ is significantly affected. Moreover, their role increases as the
proper inclination decreases. 

The right column of Fig.~\ref{fig:map_ae_big} shows the location of the resonances
$\smash{\nu_{1j}, \, j=2,3,4,6}$ involving the longitude of the node. 
The $\nu_{12}$ resonance basically never appears in the region 0.5 au $<a<$ 3 au for
inclination $i_0$ up to $40^{\circ}$. This was already noticed by
\citet{michel-froeschle_1997}, who found that $\nu_{12}$ appears only at semi-major axis
smaller than 0.5 au, at least for proper eccentricity of $0.1$. Our results confirm this feature
even at larger eccentricity values, hence this resonance should not play any
major role in the dynamics of NEOs.
The resonances $\nu_{13}$ and $\nu_{14}$ involving the Earth and Mars are again 
close to each other, since also the node of their orbits precesses at a similar rate
(see Table~\ref{tab:planet_frequencies}). 

At inclination $i_0 = 40^\circ$, $\nu_{13}$ and $\nu_{14}$ appear at $a<2$ au, while
$\nu_{16}$ is present in the region $a<2.5$ au. These three resonances all span a large
interval of eccentricities, arriving up to $e_0\approx 0.7$. Moreover, $\nu_{13}$ and $\nu_{16}$
extend towards the limit of $e_0 = 0.8$ that we used in our computations. 
As the proper inclination decreases, the secular resonance curves shrink, and they move towards
smaller values of semi-major axis and proper eccentricity. 
At $i_0 = 30^\circ$, the qualitative picture is the same as that obtained at $i_0 =
40^\circ$.
At $i_0 = 20^\circ$, the $\nu_{16}$ resonance does not change its qualitative shape. On
the other hand, $\nu_{13}$ and $\nu_{14}$ are composed by four different separated curves,
confined at eccentricity smaller than about 0.5. Similarly to the case of the $\nu_j$
resonances, the curves of constant $q, Q$ equal to the semi-major axis of a planet play a
role in determining the proper frequencies $s$, and the location of the secular
resonances.
At $i_0=10^\circ$, $\nu_{16}$ has a different qualitative structure. It shows a closed
curve in the region $a_{2} < q < a_{3}$, and another branch appears
at $q<a_{3}$. This is the only resonance appearing to a large extent, while all
the other either disappeared or are confined to a very small region of the phase space. 
At lower values of proper inclination $i_0$ we found that all the secular resonances involving the nodal 
longitude disappear.

The Kozai resonance does not involve the motion of the planets, and it occurs when
$g-s = 0$ or, equivalently, when the argument of the perihelion $\omega$ librates.
In Fig.~\ref{fig:map_ae_big} and \ref{fig:map_ae_small} the
Kozai resonance region is identified in gray.  
At $i_0 = 40^\circ$ Kozai librators are rare, and indeed the gray area is small.
At $i_0 = 30^\circ$ and $20^\circ$, the Kozai resonance appears to a large extent in
three different regions: near $a = a_{2}$ and $a= a_{3}$ at small
eccentricity, and at the intersection between $q=a_{2}$ and
$Q=a_{3}$.
At inclination $i_0 \leq 10^\circ$ only the Kozai librators near the semi-major axes of Venus and the
Earth are left, while a small gray area appear near $a = a_{4}$. Note also that
the extent of the gray area decreases as the proper inclination value decreases.

\subsection{Secular resonance maps at fixed eccentricity}
\label{ss:map_ai}
\begin{figure*}[!ht]
   \centering
   \includegraphics[width=\textwidth]{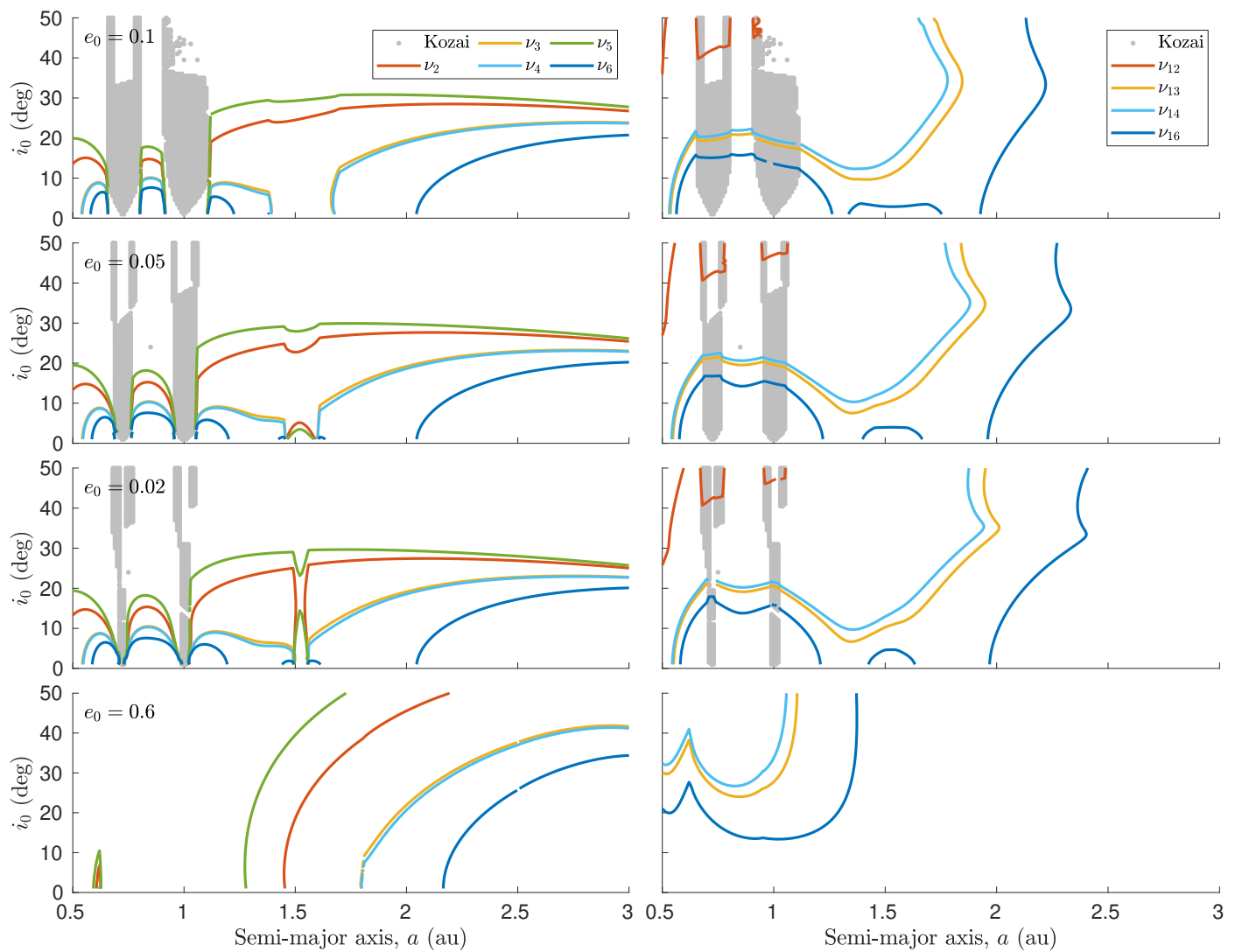}
   \caption{Map of secular resonances in the $(a,i_{0})$-plane for $a$ between 0.5 and 3 au, for different values of
      proper eccentricity $e_{0}$. The left column shows the locations of $\nu_j, j=2,3,4,5,6$, while the
      right column shows the locations of $\nu_{1j}, j=2,3,4,6$. 
      The color code of each resonance is indicated by the legend in the top row panels.
      From the top to the bottom, rows correspond to values of $e_{0}$ equal to $0.1,
      0.2, 0.4$, and $0.6$, respectively. We also reported the location of the Kozai
      librators in grey, identified by $g -s= 0$.}
   \label{fig:map_ai_big}
\end{figure*}

\begin{figure*}[!ht]
   \centering
   \includegraphics[width=\textwidth]{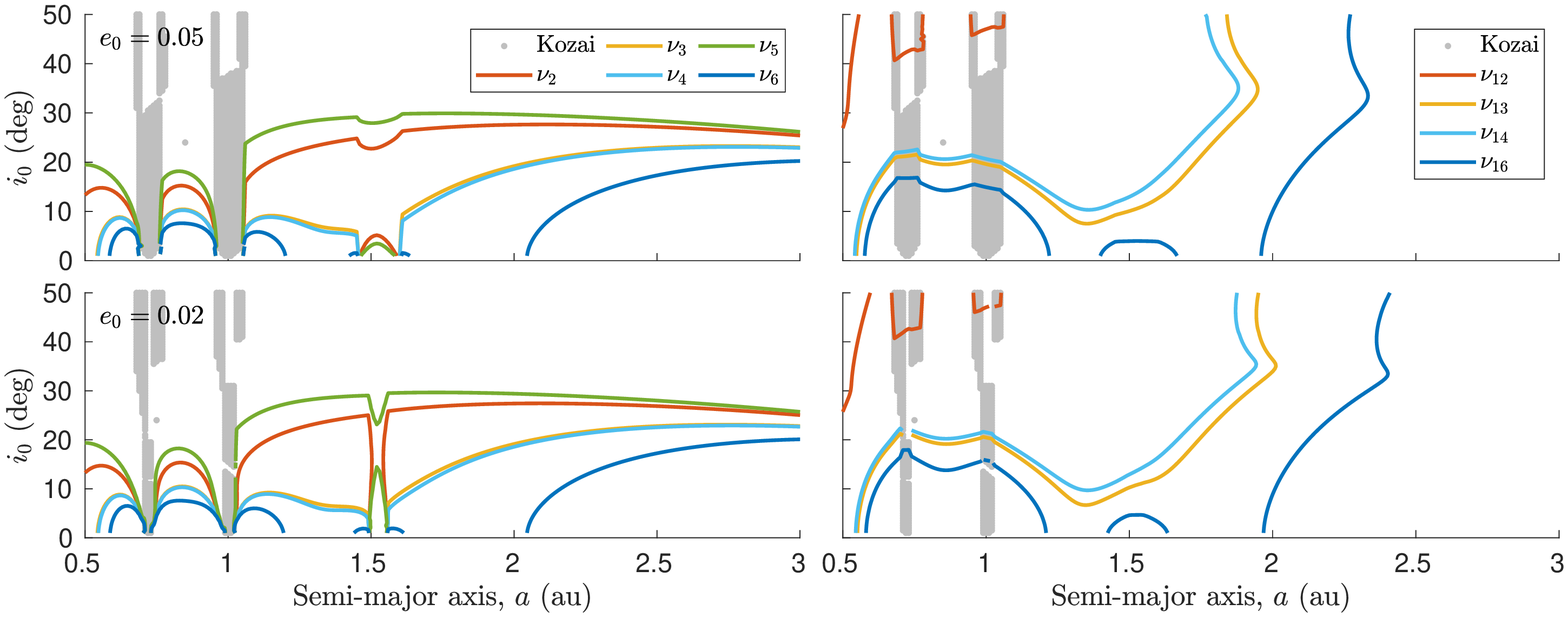}
   \caption{Map of secular resonances in the $(a,i_{0})$-plane for $a$ between 0.5 and 3 au, for different values of
      proper eccentricity $e_{0}$. The left column shows the locations of $\nu_j, j=2,3,4,5,6$, while the
      right column shows the locations of $\nu_{1j}, j=2,3,4,6$. 
      The color code of each resonance is indicated by the legend in the top-left panel.
      From the top to the bottom, rows correspond to values of $e_{0}$ equal to $0.05,
      0.02$, respectively. We also reported the location of the Kozai
      librators in grey, identified by $g -s = 0$. 
      }
   \label{fig:map_ai_small}
\end{figure*}

We discretized the $(a,i_{0})$-plane with a step of 0.01 au for the semi-major axis, and
0.5$^{\circ}$ for the inclination. We took into account the region defined by 0.5 au $<a<$ 3 au 
and $0^{\circ} < i_{0} < 50^{\circ}$. The proper eccentricity was fixed, and we computed the maps
for values $e_{0} = 0.1 \cdot h, \ h=1,\dots,7$, and for low eccentricity $e_0 = 0.02,
0.05$.
The left column of Fig.~\ref{fig:map_ai_big} shows the location of the $\nu_j, j=2,\dots,
6$ secular resonances, while the right column shows the location of $\nu_{1j}, j=2,3,4,6$, for values of
proper eccentricity $e_{0}$ equal to $0.1, 0.2, 0.4$, and $0.6$. 
Note that the first row of Fig.~\ref{fig:map_ai_big} is the same case considered by
\citet{michel-froeschle_1997}. The results we obtained are in a really good agreement with
those presented by \citet{michel-froeschle_1997} (see Fig.~2 and 3 therein), with the difference
that we are able to extend the maps to the planet-crossing regions.
Figure~\ref{fig:map_ai_small} shows the same maps obtained for $e_{0} = 0.05, 0.02$. 

From the left column of Fig.~\ref{fig:map_ai_big}, we can see that all the resonances
$\nu_j, \, j=2,\dots,6$ appear in the region 0.5 au $<a<$ 3 au. As expected, 
$\nu_3$ and $\nu_4$ are always close to each other. 
At proper eccentricity $e_0 = 0.1$, the $\nu_3, \nu_4$, and $\nu_6$ resonances all have four
different branches, each one appearing in the regions $a>a_{4}$,
$a_{3} < a < a_{4}$, $a_{2} < a < a_{3}$, and
in $a<a_{2}$. The $\nu_6$ resonance with Saturn is confined at inclination
smaller than about 20$^\circ$. The resonances $\nu_2$ and $\nu_5$ have three different
branches, each one appearing for $a>a_{3}$, $a_{2} < a <
a_{3}$, and in $a<a_{2}$. 
We can also notice the presence of Kozai librators around
$a=a_{2}$ and $a = a_{3}$, that fill this area of the phase space up to $i
\approx 35^\circ$. 

At proper eccentricity $\smash{e_0 = 0.2}$ the branches in the region $a_{2}< a <
a_{3}$ almost disappeared, while those that were appearing at
$a<a_{2}$ are shifted to smaller values of semi-major axis, if compared with
the previous case. For $a>a_{3}$, the map is qualitatively similar to the case
$e_0 = 0.1$. The Kozai librators still appear near $a_{2}$ and
$a_{3}$, but they fill a smaller area of the phase space, that is also moved
towards larger inclinations. Moreover, another region filled by Kozai librators appears
between Venus and the Earth, that corresponds to that seen in Fig.~\ref{fig:map_ae_big}
for proper inclination of $30^\circ$, and $40^\circ$. We found that the location of the
secular resonances for $e_0 = 0.3$ is overall qualitatively similar to $e_0 = 0.2$, and
therefore it is not shown.

The maps for $e_0 = 0.4$ and $e_0 = 0.6$ are also qualitatively similar to each other. The
region at $a \lesssim 1.3$ au is almost cleared out from all the secular resonances, that
appear to a large extent at larger semi-major axis value, where they only have one branch.
Moreover, they cover a larger interval of inclination values than in the previous cases.
It is also worth noting that Kozai librators disappeared, consistently with the results of
Fig.~\ref{fig:map_ae_big} and \ref{fig:map_ae_small}.

The left column of Fig.~\ref{fig:map_ai_small} shows the maps obtained for $e_0 = 0.05$
and $0.02$, and they are qualitatively similar to that obtained for $e_0 = 0.1$. We can
see that the Kozai resonance region is smaller than for the case $e_0 = 0.1$, and the
secular resonances extend closer to the semi-major axes of Venus, of the Earth, and of
Mars.

The right column of Fig.~\ref{fig:map_ai_big} shows the location of the $\nu_{1j},
j=2,3,4,6$ secular resonances. As seen in Sec.~\ref{ss:map_ae}, the resonance $\nu_{12}$
with Venus is negligible. At eccentricity $e_0 = 0.1$, $\nu_{16}$ has three different
branches, while $\nu_{13}$ and $\nu_{14}$ have only one. These three resonances are
confined at semi-major axis smaller than $2.3$ au, and they span a large interval of
inclination values.
Note that $\nu_{13}, \nu_{14}$, and $\nu_{16}$ appear also in the Kozai resonance region. 
For larger value of proper eccentricity, $\nu_{16}$ also have a single branch. Moreover,
as $e_0$ increases, their location move towards smaller values of semi-major axis, and
they span a smaller interval of inclinations. 
Finally, the right column of Fig.~\ref{fig:map_ai_small} shows the location of the $\nu_{1j}$ resonances for $e_0 = 0.02$ and $0.05$. The maps are qualitatively similar to the case $e_0 = 0.1$, with the difference that the three resonances involving the Earth, Mars, and Saturn all extend towards larger semi-major axis values.

\subsection{Frequencies near perihelion equal to planets' semi-major axes}
\begin{figure}[!ht]
    \centering
    \includegraphics[width=0.48\textwidth]{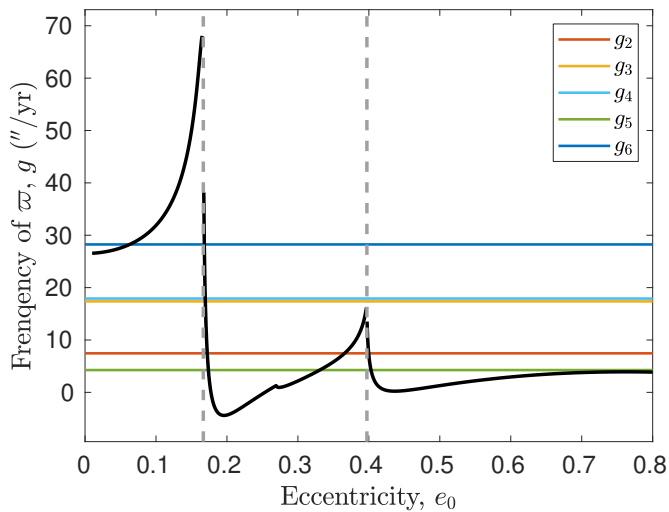}
    \caption{Frequency $g$ (black solid curve) computed on a grid in proper eccentricity $e_0$, for a fixed value
    of semi-major axis $a=1.2$ au and proper inclination $i_0 = 2^\circ$. The horizontal straight lines correspond to the values $g_j, j=2,3,4,5$ of the proper frequencies of the planets. The vertical dashed lines correspond to the values of eccentricity such that $a(1-e_0) = a_j, j=2,3$.}
    \label{fig:g_slice}
\end{figure}

Figures~\ref{fig:map_ae_big}, \ref{fig:map_ae_small} show that the secular resonances tend to follow the perihelion curves $a(1-e) = a_j, j=2,3,4$, and the trend is more prominent when the proper inclination $i_0$ decreases towards 0. To better understand the reason of this, we computed the proper frequencies on a grid in eccentricity, by fixing both the values of the semi-major axis and of the proper inclination. To have a better picture of the behaviour of the frequency, we also used a step in eccentricity of $5\times10^{-4}$, that is smaller than that used to produce Fig.~\ref{fig:map_ae_big} and \ref{fig:map_ae_small}. Figure~\ref{fig:g_slice} shows the frequency $g$ computed for $a=1.2$ au and $i_0 = 2^\circ$, as a function of the proper eccentricity $e_0$. When $e_0$ approaches the critical values $e_0^{(j)}, j=2,3$ for which $a(1-e^{(j)}_0) = a_j, j=2,3$, the frequency $g$ sharply increases. Moreover, the magnitude of the increase is larger and larger as the inclination tends to $0^\circ$. This feature is caused by the model itself, and it is due to the orbit crossing singularity.
\begin{figure}[!ht]
    \centering
    \includegraphics[width=0.48\textwidth]{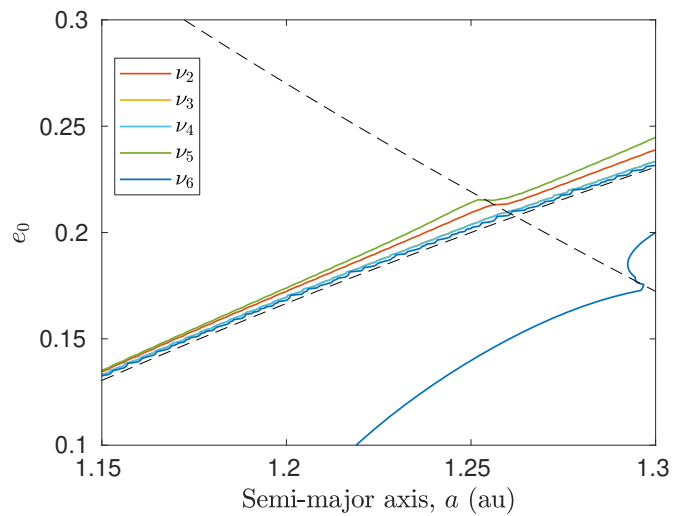}
    \caption{Zoom of the map of the secular resonances $\nu_j, j=2,3,4,5,6$ at $i_0 = 2^\circ$, near $a=1.2$ au and $e_0 = 0.17$. }
    \label{fig:g_zoom}
\end{figure}

Figure~\ref{fig:g_slice} shows also an interesting feature. Between $\smash{e_0^{(3)}\approx 0.1667}$ and $\smash{e_0^{(2)} \approx 0.3972}$ the frequency $g$ has a minimum that is smaller than $g_5$, that is the smallest planetary frequency. As $e_0$ tends to $e_0^{(3)}$ from the right, the frequency $g$ increases sharply, and it crosses all the planetary frequencies $\smash{g_j, j=2,\dots,6}$ in a small interval of proper eccentricities $e_0$. The same happens as $e_0$ tends to $e_0^{(2)}$ from the right, with the difference that only $g_2$ and $g_5$ are attained near the critical value. 
This effect causes the secular resonances to follow the Earth perihelion curve, and to be close to each other. In addition, the fact that $g$ grows so fast in such a small interval of eccentricity will make the resonance width very small, so that objects following the perihelion curves by effect of the secular resonances will be extremely rare to find, because they would be pushed outside of the resonance by other small perturbations. In fact, we did not find any object following a planet's perihelion curve in our purely numerical integrations (see Sec.~\ref{ss:numericalIntegrations}).

The sharp increase of the frequency $g$ near planets' perihelion curves causes also some small numerical errors in the global plot of the maps presented in Fig.~\ref{fig:map_ae_big} and \ref{fig:map_ae_small}. Because we use a discrete grid in the $(a,e_0)$-plane, it may happen that a value larger than one of the planets' frequency $g_j$ is not achieved in the neighbourhood of one of such perihelion curves. As a consequence, the corresponding secular frequency $\nu_j$ would not appear in the plot. This can be seen for instance in the maps at $i_0=2^\circ$, presented in Fig.~\ref{fig:map_ae_small}. Figure~\ref{fig:g_slice} shows that all the five resonances $\nu_j, j=2,\dots,6$ are crossed in a neighbourhood of $a=1.2$ au and $e_0 \approx 0.17$, while this is not clear to happen in the left panel shown in Fig.~\ref{fig:map_ae_small}. This is caused by the fact that the grid is not dense enough in this tiny neighbourhood. To prove this, we recomputed the secular frequencies in the portion $(a, e_0) \in [1.15 \text{ au}, 1.3 \text{ au}] \times [0.1, 0.3]$, using a step of 0.004 au in semi-major axis and of $5\times 10^{-4}$ in proper eccentricity. Figure~\ref{fig:g_zoom} shows the location of the secular resonances in this portion of the phase space. Here we can see that all the five secular resonances $\nu_j$ appear close to $q=a_3$, even though $\nu_3, \nu_4$ and $\nu_6$ are barely distinguishable since they are very close to each other. Overall, we found these numerical effects to happen at low proper inclination $i_0$ and only near $q=a_j, j=2,3,4$, where the width of the resonance is very small in any case. Therefore, using a denser grid would not change the global picture of the secular resonances shown in Sec.~\ref{ss:map_ae}, that was obtained using a less dense grid.

\subsection{Comparison with numerical integrations}
\label{ss:numericalIntegrations}
We performed purely numerical integrations to confirm the secular resonance locations 
predicted by the theory in Sec.~\ref{ss:map_ae} and \ref{ss:map_ai}. 
To account for the fact that the maps are computed in the proper elements space, we
took initial orbital elements along the curves $g = g_j$ (or $s = s_j$) and $g = g_j \pm
1^{\prime\prime}/\text{yr}$ (or $s = s_j\pm 1^{\prime\prime}/\text{yr}$), for a selected
planet $j \in \{ 2,3,4,5,6 \}$. The initial argument of pericenter $\omega$ and the
initial argument of the node $\Omega$ were both set to $0^\circ$, while the initial mean
anomaly $\ell$ was chosen randomly between $0^\circ$ and $360^\circ$.
Numerical integrations were performed by using a hybrid symplectic scheme
\citep{chambers_1999} that is able to handle close encounters with planets, that is included in
the
\textsc{mercury}\footnote{\url{https://gemelli.spacescience.org/~hahnjm/software.html},  \url{https://github.com/Fenu24/mercury}}
package \citep[see also][]{fenucci-novakovic_2022}. 
The gravitational attraction of the Sun and of all the planets from Mercury to Neptune were
included in the model, and the initial conditions for planets at epoch 58800 MJD were
taken from the JPL Horizons\footnote{\url{https://ssd.jpl.nasa.gov/?horizons}} ephemeris
system. Note that here we take into account a complete model of the Solar System to show that the simplified model presented in Sec.~\ref{s:methods} is actually able to predict the location of secular resonances in a qualitative picture.
We propagated the evolution of the asteroids and the planets for a total time of 2 Myr, by
using a timestep of 1 day. 

\begin{figure}[!ht]
    \centering
    \includegraphics[width=0.48\textwidth]{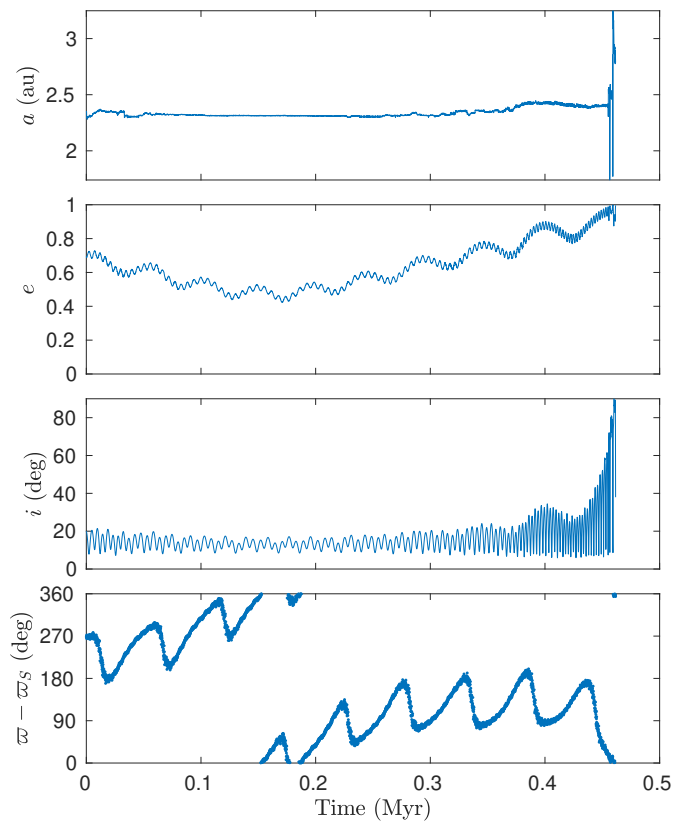}
    \caption{Evolution of a test particle inside the $\nu_6$ secular resonance. The first three panels show the evolution of semi-major axis, eccentricity, and inclination, while the last panel shows the evolution of the critical angle $\varpi - \varpi_S$.}
    \label{fig:nu6_20}
\end{figure}

We present here the results obtained by choosing an initial inclination of $20^\circ$.
Figure~\ref{fig:nu6_20} shows the evolution of a test asteroid selected with the method
described above, that is affected by the $\nu_6$ secular resonance. The critical angle
$\varpi-\varpi_S$ librates around 0$^\circ$, while the eccentricity decreases down to
values of about 0.4 during the first 0.2 Myr of evolution, and then suddenly increases up
to values close to 1, causing this object to impact the Sun in less than 0.5 Myr
evolution. This is a well known effect of the $\nu_6$ resonance, found for the first time
by \citet{farinella-etal_1994}, that is able to deliver Sun impactors even starting from
the main belt at very low eccentricity. This simulation shows that, as expected, this
behaviour still occurs well inside the NEO region, even at higher values of the initial
eccentricity.

\begin{figure}[!ht]
    \centering
    \includegraphics[width=0.48\textwidth]{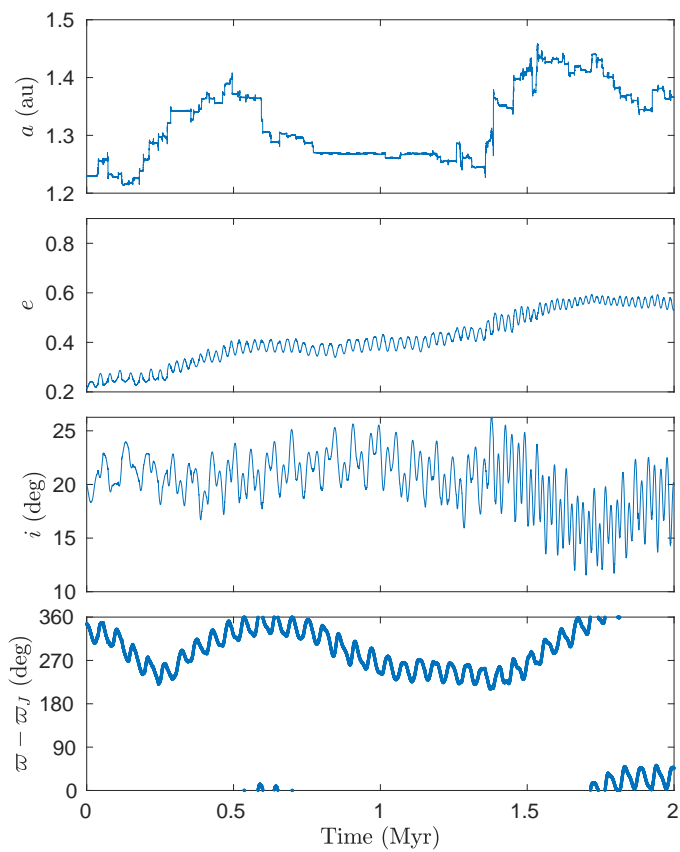}
    \caption{Same as Fig.~\ref{fig:nu6_20}, for a test particle in the $\nu_5$ secular resonance. The last panel shows the evolution of the critical argument $\varpi-\varpi_J$.}
    \label{fig:nu5_20}
\end{figure}

Figure~\ref{fig:nu5_20} shows a test asteroid affected by the $\nu_5$ secular resonance.
The critical angle $\varpi-\varpi_J$ librates around 0$^\circ$ during the 2 Myr of evolution,
and the eccentricity passes from an initial value of about 0.2 to values close to 0.6.
The semi-major axis assumes values between 1.2 and 1.5 au, and it undergoes a random walk
evolution, caused by close encounters with the inner planets. 
Despite the changes in the
semi-major axis, the critical angle still librates and it is not removed from the secular
resonance.
Although this object did not end the evolution on a collision with the Sun, $\nu_5$ was
able to push it to large eccentricity values, showing a behaviour similar to that of
$\nu_6$.  By using numerical simulations, \citep{gladman-etal_2000} showed that a
combination of planetary close encounters and the $\nu_5$ secular resonance is a route to
Sun-grazing orbits at semi-major axis $a<2$ au. Therefore, $\nu_5$ could provide another
significant mechanism for the delivery of asteroids impacting the Sun and the inner planets,
that acts well inside the NEO region.

\begin{figure}[!ht]
    \centering
    \includegraphics[width=0.48\textwidth]{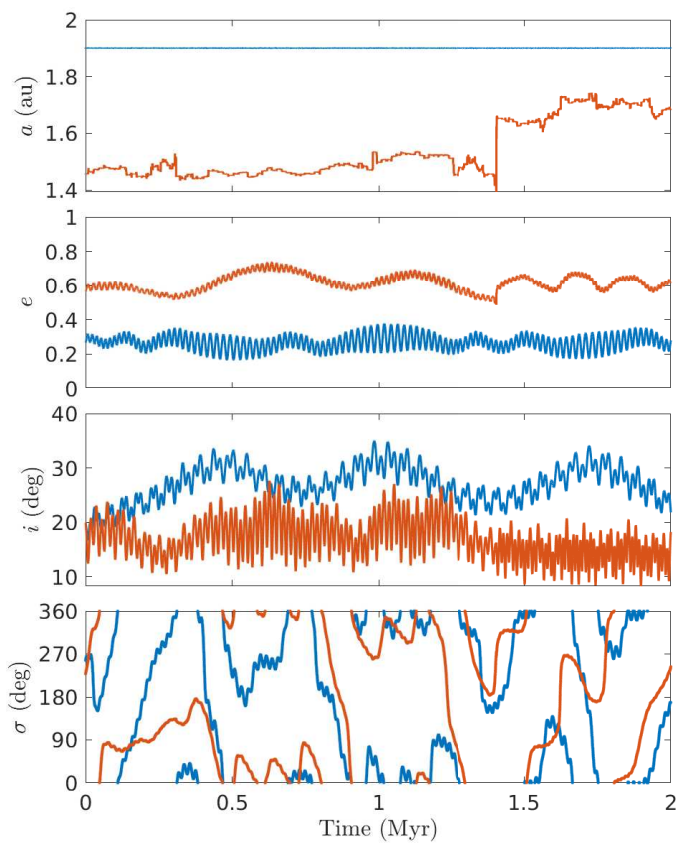}
    \caption{Same as Fig.~\ref{fig:nu6_20}, for two test particles in the $\nu_2$ (red
 curve) and the $\nu_3$ (blue curve) secular resonances. The last panel shows the evolution
 of the corresponding critical argument $\sigma$: $\varpi-\varpi_V$ (red curve) for
 $\nu_2$, and $\varpi-\varpi_E$ (blue curve) for $\nu_3$.}
    \label{fig:nu2_3_20}
\end{figure}

Figure~\ref{fig:nu2_3_20} shows two test asteroids located near $\nu_2$ and $\nu_3$.
These objects both show the corresponding critical argument librating around
0$^\circ$ but, contrary to the previous examples, these resonances do not increase the
eccentricity in the 2 My integration time-span. On the other hand, the test asteroid affected by $\nu_3$ shows an initial
increasing of the inclination up to $30^\circ$ caused by the effect of the resonance.

Here we showed only few examples corresponding to one of the maps of
Fig.~\ref{fig:map_ae_big}, however all the numerical results obtained are in agreement with
the locations computed for $i_0 = 20^\circ$.  Therefore, we think that the maps obtained in
Sec~\ref{ss:map_ae} and \ref{ss:map_ai} provide a reasonable estimation for the location
of the secular resonances in the NEO region.

\section{Discussion}
\label{s:discussion}

\subsection{Possible dynamical paths inside the NEO region}
\begin{figure*}[!ht]
   \centering
   \includegraphics[width=\textwidth]{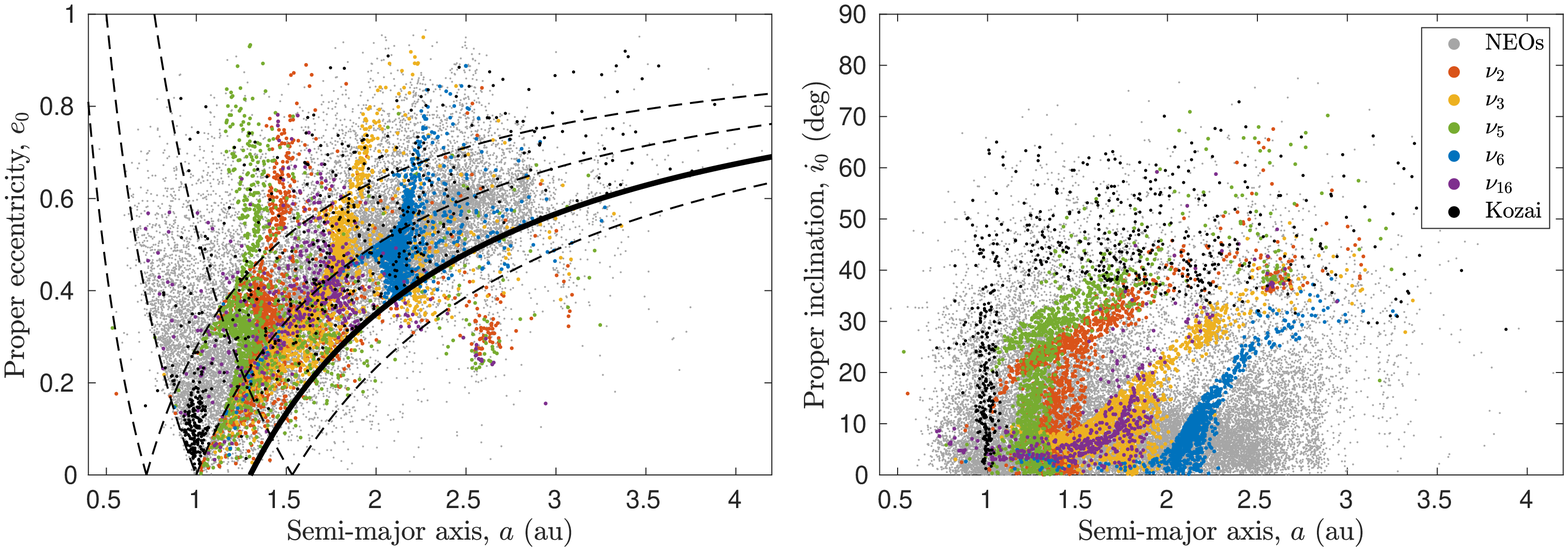}
   \caption{Known NEOs close to some secular resonances, in the phase space $(a, e_0)$
      (left panel) and $(a, i_0)$ (right panel). Gray dots represent the background
      population of NEOs, while red, yellow, green, blue, and violet
      correspond to the $\nu_2, \nu_3, \nu_5, \nu_6$ and $\nu_{16}$, respectively. Black
      dots denote objects in the Kozai resonance.}
   \label{fig:knownNEOs}
\end{figure*}
\citet{farinella-etal_1994} showed that the $\nu_6$ secular resonance is very 
efficient in increasing the eccentricity of asteroids initially located in the main belt, causing $e$
to pass from values near to 0 to values near to 1 in just $\sim$0.5 Myr. 
\citet{foschini-etal_2000, gladman-etal_2000} showed some numerical example in which the
$\nu_5$ secular resonance increases the eccentricity of NEOs located at $a < 2$ au,
causing asteroids to end up in a collision with the Sun. However, the location of the $\nu_5$ resonance inside the
NEO region was not determined. 
In Sec.~\ref{ss:numericalIntegrations}, we also showed an example where $\nu_5$ 
is able to pump up the eccentricity to values near to 1. 
Thus, some secular resonances are responsible for the delivery of asteroids impacting the
planets or the Sun, even well inside the NEO region.

To search for possible dynamical paths, we took the population of known NEOs and identified
those that are near the secular resonances. 
The nominal orbital elements were downloaded from the
Near-Earth Objects Coordination Centre\footnote{\url{https://neo.ssa.esa.int/}} (NEOCC) on
24$^{\text{th}}$ August 2022. Then, we computed their proper frequencies $g, s$, and their proper
elements $e_0, i_0$ as we defined in Sec.~\ref{s:methods}. We plotted all the NEOs in the
$(a,e_0)$ and in the $(a,i_0)$ planes, and highlighted the objects near the
$\nu_{j}, j=2,3,5,6$ and $\nu_{16}$ secular resonances, as well as Kozai librators. 
NEOs near $\nu_4$ were not plotted, because they almost overlap with
$\nu_3$. Since we did not attempt to compute the width of the resonances, 
we set an arbitrary threshold of 1$^{\prime\prime}$/yr for the identification of resonant
objects, that we
believe to be a reasonable choice \citep[see also][]{gronchi-milani_2001}. 
Figure~\ref{fig:knownNEOs} shows the result of the plot.
Note that secular resonances appear as a swarm of points in this figure, differently to what shown in Sec.~\ref{s:results}. This is because here we show the projections of the whole 3-dimensional space of the proper elements on a 2-dimensional plane, and also because we assumed a non-zero resonance width.
From the plot in the $(a,e_0)$-plane, we can notice that objects near to $\nu_6$ cross the
curve $q = a_{3}$, confirming again that this resonance is able to move objects onto
Earth-crossing orbits \citep{bottke-etal_2002, granvik-etal_2018}.
Moreover, NEOs near $\nu_j, j=2,\dots,6$ extend up to large eccentricity values, even over 0.8. 
The fact that known NEOs are currently near these resonances, and are placed at high
eccentricity, is an indication that all the $\nu_j$ resonances may be able to pump up the eccentricity and
deliver objects impacting the Sun.
It is worth also noting that there are NEOs near $\nu_{16}$ between 1.5 au and 2 au, that
extend from the border of the NEO region at $q=1.3$ au to $q=a_{3}$. Therefore,
also this resonance could be able to move asteroids onto Earth-crossing orbits. Note that
$\nu_{16}$ is near to the Hungaria region \citep[see e.g.][]{froeschle-scholl_1986}, that
is now recognized to be an important source region of NEOs \citep{granvik-etal_2018}
producing a non-negligible fraction of Earth impactors.

\citet{wetherill_1988} suggested that the $\nu_{16}$ secular resonance could be
responsible for the production of NEOs with inclination larger than $30^{\circ}$.
Later, \citet{gladman-etal_2000} showed with numerical simulations that this mechanism is
actually possible. In the right panel of Fig.~\ref{fig:knownNEOs}, we can see objects at high
inclinations near $\nu_2, \nu_5$, and $\nu_{16}$, suggesting that they all could be
capable of increasing the inclination. Indeed, the maps of Fig.~\ref{fig:map_ai_big} show
that $\nu_2$ and $\nu_5$ extend towards values of the inclination larger than
50$^{\circ}$ for proper eccentricity of 0.6. Additionally, $\nu_{12}, \nu_{13}$, and
$\nu_{16}$ all extend towards high inclination values, regardless of the value of the
proper eccentricity. Therefore, even if not reported in Fig.~\ref{fig:knownNEOs}, also $\nu_{12}$ and
$\nu_{13}$ could be responsible for increasing the inclination to values higher than
30$^{\circ}$.

In this work, we only gave an indication of possible dynamical paths inside the NEO region, 
that are based on the secular motion and on the current position of known NEOs. 
Additional extensive numerical simulations with a full $N$-body model need to be performed 
to better investigate and establish these hypotheses.

\subsection{Effects of mean-motion resonances}
The interaction between secular resonances and mean-motion resonances (MMRs) with
Jupiter has been studied by \citet{morbidelli-moons_1993, moons-morbidelli_1995} in the
main belt.
The results presented in this paper were obtained with a semi-analytical model that does
not take into account the effect of MMRs between the asteroid and a planet.
A method for the computation of proper elements and frequencies of resonant NEOs, that
extends the one by \citet{gronchi-milani_2001}, has been recently developed by
\citet{fenucci-etal_2022}.
The authors showed that the secular evolution of asteroids that are placed well
inside the strongest MMRs with Jupiter is significantly different from
the non-resonant secular evolution, and the frequencies $g$ and $s$ are generally
affected by these resonances. The same happens for some MMRs with the Earth, and Venus. On
the contrary, MMRs with Mars have generally little effect on the
long-term dynamics.

Some of the strongest MMRs with Jupiter, such as the
5:2, 7:3, and 2:1, are located beyond 2.5 au. The 3:1 MMR, that is the strongest
one with Jupiter, is located at about 2.5 au. 
Therefore, they should not modify the results obtained for the nodal resonances $\nu_{1j}$, 
since they always appear at $a<2.5$ au, while they may somewhat affect the location of
$\nu_3, \nu_4,$ and $\nu_6$ at inclinations $i_0$ larger than about 20$^\circ$.
On the other hand, the 7:2, 4:1, and 5:1 MMRs with Jupiter occur at 
smaller semi-major axis values, i.e. at about 2.25 au, 2.06 au, and 1.78 au respectively.
Additionally, low-order resonances with Venus and the Earth mostly appear at $a<2.5$ au
\citep[see e.g.][]{gallardo_2006}, hence they could all somewhat affect the secular
resonances location. 

Recently, \citet{zhou-etal_2019, xu-etal_2022} determined maps of secular resonances near the 1:1 MMR with the Earth and with Venus, computing the proper frequencies by using pure numerical integrations and frequency analysis \citep{laskar_1988, laskar_1990, laskar_2005}. The maps obtained in these works do show some differences with respect to those presented in this paper. However, the effects of MMRs is only local, and the secular resonances are affected only in a strip in semi-major axis of $\sim$0.02 au width for the 1:1 MMR with the Earth, and of $\sim$0.008 au width for the 1:1 MMR with Venus. 

\begin{figure}[!ht]
   \centering
   \includegraphics[width=0.48\textwidth]{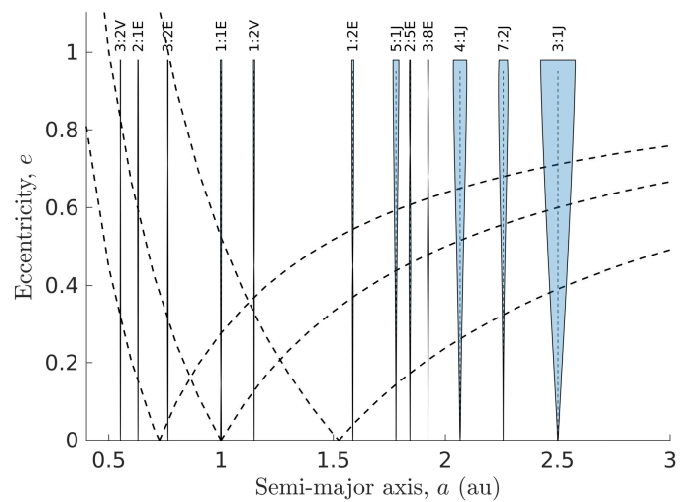}
   \caption{Strongest mean-motion resonances at semi-major axis $a<2.5$ au. The filled light blue
      area represents the resonance width, computed with an analytic method
      \citep[see][]{morbidelli_2002} for a fixed value of inclination of $15^\circ$.}
   \label{fig:mmrwidth}
\end{figure}
To give a bigger picture of the strongest MMRs appearing at $a<2.5$ au, we show their width if 
Fig.~\ref{fig:mmrwidth}. The width was computed with a simplified analytical
method taken from \citet{morbidelli_2002}, for a fixed value of inclination of 15$^\circ$. 
MMRs generally occupy a small volume of this part of the phase space, and those with Venus and the Earth 
are especially narrow. This observation, coupled with the fact that MMRs significantly change the secular
dynamics only when the asteroid is well inside the resonance \citep{fenucci-etal_2022}, should limit the 
effect of MMRs on the results presented in this paper, that are aimed to give the general picture of the whole NEO region.

\section{Conclusions}
\label{s:conclusions}
In this paper, we determined the location of the secular resonances with the planets from Venus to Neptune inside the NEO region. Proper elements and proper frequencies of NEOs were computed by using a semi-analytical secular model that permits to propagate the secular evolution of objects
beyond orbit-crossing singularities. 
Maps of the secular resonances were then obtained by keeping the proper inclination $i_0$
(or the proper eccentricity $e_0$) at a fixed value, and we showed how the locations change by
varying the value of the fixed proper element. 

The $\nu_j, j=2,\dots,6$ secular resonances all appear well inside the NEO region,
especially at inclinations smaller than about $30^\circ$. On the other hand, the
$\nu_{1j}$ secular resonances with the node of the planets appear to a large extent
only for $j=3,4,6$, and they tend to be negligible at small inclination. 
To confirm the computed locations, we performed full numerical $N$-body simulations that
included all the planets of the Solar System. The results obtained showed that the maps
computed with the semi-analytical approach provide a good estimate of the secular
resonance locations. 
The current distribution of NEOs close to secular resonances suggests that some of these
resonances could be able to produce Sun-grazing asteroids and Earth impactors, or they may
increase the inclination to values larger than 30$^\circ$.


\begin{acknowledgements}
The authors have been supported by the MSCA-ITN Stardust-R, Grant Agreement 
n. 813644 under the European Union H2020 research and innovation program. 
GFG also acknowledge the project MIUR-PRIN 20178CJA2B
“New frontiers of Celestial Mechanics: theory and applications" and the
GNFM-INdAM (Gruppo Nazionale per la Fisica Matematica).
\end{acknowledgements}

\bibliographystyle{aa}
\bibliography{holybib.bib}{} 

\end{document}